\documentclass[prl,aps,tightenlines,twocolumn,showpacs]{revtex4}
\usepackage{epsfig}

\begin{document}
\title{Decoherence of geometric phase gates}

\author{A. Nazir\footnote{Present address: 
Department of Materials, University of Oxford, OX1 3PH, UK.}, T. P. Spiller and W. J. Munro}

\affiliation{Hewlett-Packard Laboratories, Filton Road,
Stoke Gifford, Bristol BS34 8QZ, UK}

\date{\today}

\begin{abstract}
We consider the effects of certain forms of decoherence applied to both
adiabatic and non-adiabatic geometric phase quantum gates. For a single 
qubit we illustrate
path-dependent sensitivity to anisotropic noise and for two qubits
we quantify the loss of entanglement as a function of decoherence.
\end{abstract}

\maketitle
The discoveries of quantum algorithms for factorization \cite{shor1} and 
searching \cite{grov1}, and techniques for quantum error correction 
\cite{shor2,steane1} (and subsequent fault-tolerant methods) have generated
considerable motivation for the realisation of quantum computing (QC) hardware.
Significant progress has been made at the few qubit level; at present
many possible alternative routes are under exploration \cite{fortschritte}.
The use of fundamental entities as qubits currently leads the way; however,
there is growing longer term hope that fabricated condensed matter systems
(maybe still using fundamental rather than fabricated qubits) may
provide the vehicle for scalability in qubit number. Nevertheless, at
present the decoherence problems in such systems loom very large indeed.
Fault-tolerant techniques cannot be brought into play unless the 
underlying decoherence rates are small in the first place.

As arbitrary single-qubit gates and some entangling two-qubit gate are
universal for quantum computing \cite{div1,lloyd1,barenco1}, experimental
QC focuses on realising such gates. Whilst it would be
unfair to deem single-qubit gates trivial, the entangling of qubits 
represents the first
major hurdle---one cannot claim to have a serious QC candidate
until this has been achieved. Nevertheless, with any new contender it is
natural to investigate the simplest gates first, before progressing to those 
based on
qubit coupling. Clearly a method for inferring the likely level of
entanglement in a two-qubit gate from the results of single qubit
experiments is a handy tool. This is the basis of the simulation
approach we present here in the form of a specific example. It is certainly
true that detailed calculations of decoherence effects (based on
tracing over microscopic environments) can yield considerable
understanding of the {\em form} of decoherence seen by a qubit, and can
sometimes give estimates of decoherence rates; however, in any 
experimental realisation the
ultimate calibration of decoherence is through measurement. Given this, we
consider a simple simulation approach which, based on the observations of
single qubit gates, enables estimation of the level of entanglement (if any!)
that could be expected in a two-qubit gate with similar qubits. 
This relies on some advanced knowledge of the form of the dominant
decohering mechanisms (although some of this may be inferred from various
independent
single-qubit experiments) and an assumption of environment effects
in interacting two-qubit experiments being similar to those in the
single-qubit
experiments. If the former are worse, for example due to additional
external source terms in
the Hamiltonian, then an optimistic upper bound on entanglement results.
Despite these constraints, the simulation approach could be very useful for
new experimental QC investigations.

To illustrate the approach, we apply it to geometric phase gates
\cite{jones1,ekert1,falci1}. This provides an example of the approach; however,
the specific case of geometric phase gates is of interest in its own right,
since the technique has already been applied to NMR experiments \cite{jones1},
has been proposed for use with superconducting qubits \cite{falci1} and, in
principle, can be applied to other realisations of qubits with suitable
source terms in their Hamiltonians. We quantify the effects of different
forms of decoherence applied to single qubit phase gates, and the loss of
entanglement for a conditional two-qubit gate. The same approach can also be
applied to all forms of dynamically generated gates; a broader detailed study
will be presented in a future paper.

The model of decoherence we use is Markovian, with the reduced density operator
$\rho$ of the qubit system
described by a Bloch-type master equation
\begin{equation}
\dot \rho = -i \left[ H,\rho \right] +\sum_m\left( L_m\rho
L_m^{\dagger }-\frac 12L_m^{\dagger }L_m\rho -\frac 12\rho L_m^{\dagger
}L_m\right) \;, 
\label{eq1}
\end{equation}
where $H$ is the system Hamiltonian, the operators $\{L_m\}$ represent the 
coupling to the
environment and $\hbar = 1$. The implicit origin of the non-unitary
evolution generated by the $\{L_m\}$ is coupling to a bath
of environment degrees of freedom,  which are traced out to give the
reduced $\rho$. The form of Eq.(\ref{eq1}) is somewhat restrictive, but
within this Markovian limit it is possible to describe phenomena such as
dissipation (spontaneous decay and finite temperature stimulated effects),
white noise Hamiltonian terms and quantum measurement interactions. It is
therefore possible to treat a number of realistic forms of decoherence.
The master equation (\ref{eq1}) can be solved in many simple cases; however,
in order to be able to treat the types of time dependent Hamiltonians
(including pulses etc.) used for the realisation of actual quantum
gates, we generally work numerically. We use a quantum trajectory method,
quantum state diffusion \cite{perc1,schack1,schack2}, to solve the master
equation (\ref{eq1}) through averages over stochastically evolving quantum
states. Whilst we only give statistical data here, it is worth noting that
(ensemble NMR systems aside) since actual quantum gates/computations run on
individual systems, such simulation techniques in fact produce results in a
manner very akin to actual QC experiments and examination of individual
trajectories can provide additional insight \cite{perc1}. In all examples
presented here the averages are over 1000 trajectories unless otherwise stated.

Our first example is a single qubit geometric phase gate. In order to
generate a purely geometric phase, the dynamical phases acquired by the
different amplitudes in an arbitrary qubit state have to be cancelled
\cite{ekert1,falci1}. We use the scenario of Ref.10. The
path traversed in parameter space is amenable to the investigation of
different forms of decoherence and it has also been implemented experimentally
\cite{jones1}. A spin qubit ($|\uparrow_{z}\rangle \equiv |0\rangle$ ,
$|\downarrow_{z}\rangle \equiv |1\rangle$) is subject to a static $z$-magnetic
field $\omega_{0}$ and (within the usual rotating wave approximation) a
field of amplitude $\omega_{1}$ and phase $\phi$ at angular frequency $\omega$.
The general Hamiltonian is
\begin{equation}
H = \frac{\omega_{0}}{2} \sigma_{z} + \frac{\omega_{1}}{2}\left[
\cos(\omega t + \phi) \sigma_{x} + \sin(\omega t + \phi) \sigma_{y} \right],
\label{eq2}
\end{equation}
and to effect an ideal phase gate the spin is subjected to the unitary
sequence
$U_{4 \gamma} = \Pi \, \bar{T} \,\bar{C} \, T \, \Pi \, \bar{T} \,C  \,T$ .
Here $T$ is a tipping of the magnetic field through angle $\theta$ (ramping
$\omega_{1}$ from zero, with $\cos \theta =
(\omega_{0} - \omega)/\sqrt{(\omega_{0} - \omega)^{2} + \omega_{1}^{2}}$
with $\phi$ at zero,
$C$ is a $2\pi$ rotation of the phase $\phi$ at fixed $\omega_{1}$ and the
bars denote the reversed paths. These operations have to be carried out
adiabatically to avoid errors in the spin component
amplitudes \footnote{In principle the central $T$ and $\bar{T}$
can be omitted if $\Pi$ is done by a $\pi$ rotation about the $y$-axis
(and we do so), but
these central tips were employed in the experiments
reported in \cite{jones1}.}. Fast
$\pi$-pulses $\Pi$ interchange the $\uparrow$ and $\downarrow$ amplitudes
half way through and at the end (to cancel the dynamical phase contributions).
The ideal gate to effect a relative phase of $\gamma_{B} = 4 \gamma$ on
$|\psi\rangle \equiv 2^{-1/2} (|0\rangle + |1\rangle)$ is
\begin{equation}
U_{4 \gamma} |\psi\rangle = 2^{-1/2}
(\exp(-2 i \gamma) |0\rangle + \exp(2 i \gamma) |1\rangle),
\label{eq3}
\end{equation}
where $\gamma =
\pi(1 - \cos \theta)$ is the solid angle subtended by $C$ at the origin.
Part of the appeal of this form of quantum gate is its potentially different
sensitivities to different forms of decoherence; a potential drawback is the
need for adiabaticity, so the gate is slow and decoherence has longer to bite.

We have studied these effects in detail. Results are shown in Fig. 1(a) for the
effects of noise in the $x$- or $z$-components of the magnetic field, for two
different $\gamma_{B}$. The gate was run adiabatically giving a zero
decoherence fidelity \footnote{In terms of measurable quantities the gate
fidelity is
$f = 
\frac{1}{2}\left[1 - \sin \theta \cos \gamma_{B} \langle \sigma_{z} \rangle +
\cos \theta \cos \gamma_{B} \langle \sigma_{x} \rangle +
\sin \gamma_{B} \langle \sigma_{y} \rangle\right]$.} of
$f=0.999984$ for $\gamma_B=\pi$ and
$f=0.999993$ for $\gamma_B=\pi/8$. For the smaller
$\gamma_{B}$, which corresponds to a tipping of $14.3615\deg$ 
(so the instantaneous
energy eigenstates remain closer to $\uparrow$ and $\downarrow$ in $z$),
the system is clearly significantly more sensitive to $z$-noise compared to
$x$-noise \footnote{In effect, there is more gate
sensitivity to random fluctuations
in the energy eigenvalues (the eigenstates stay close to those of
$\sigma_{z}$) than to random precessions about the $x$-axis.}.
On the other hand, for the larger $\gamma_{B}$ where the tipping is 
$41.4096\deg$ there
is less distinction.
\begin{figure}[!h]
\center{\epsfig{figure=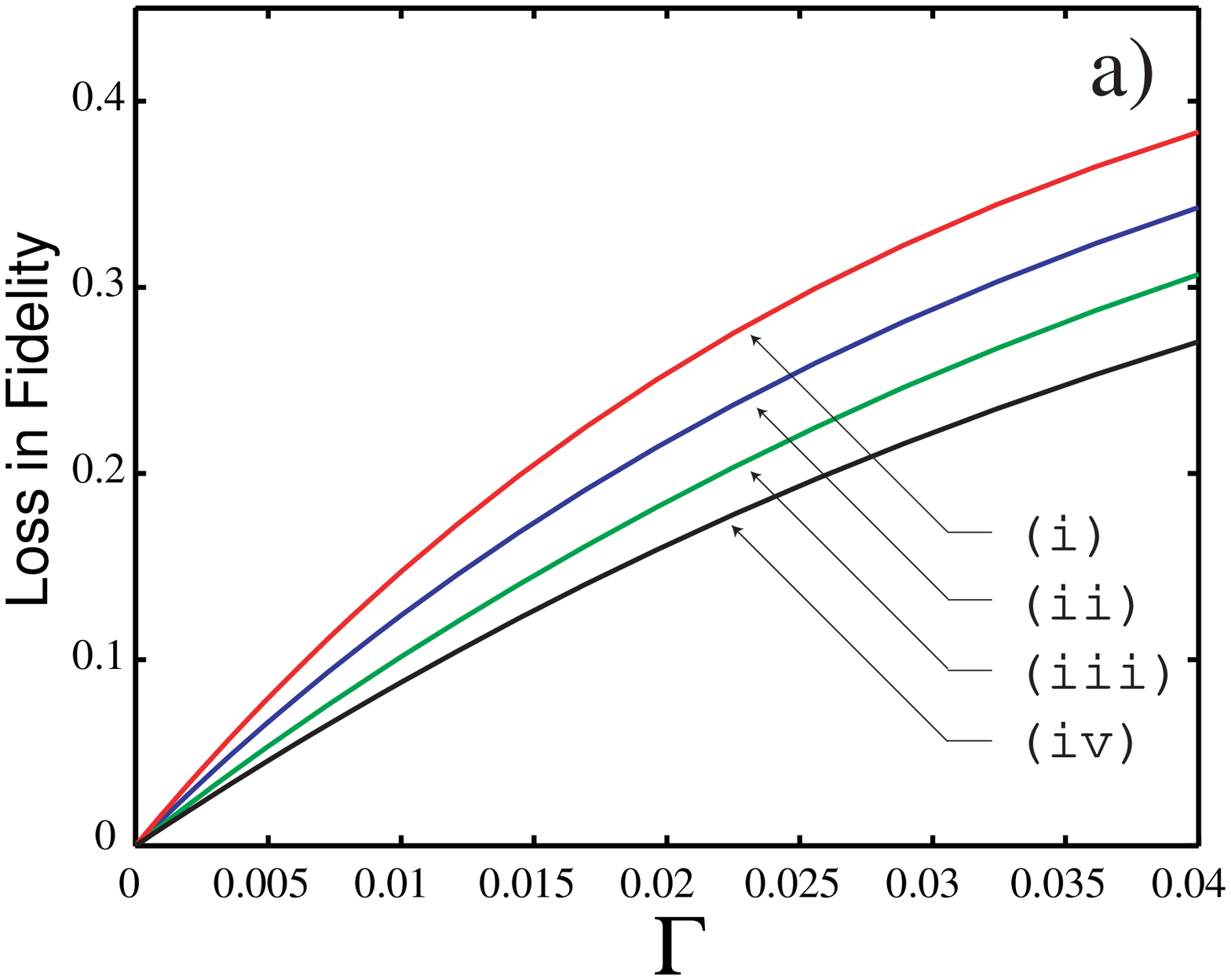,width=40mm}$\;\;\;$
\epsfig{figure=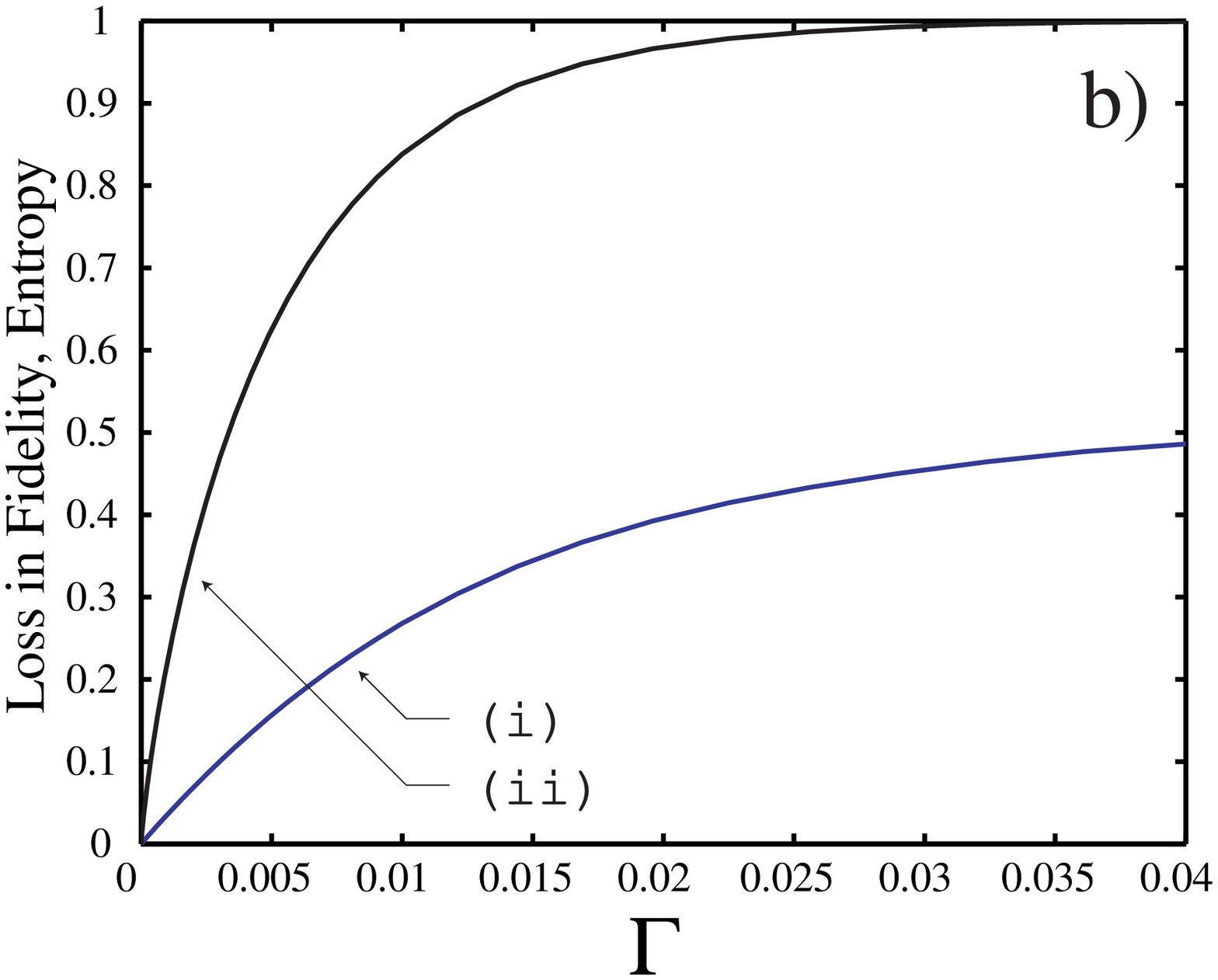,width=40mm}}
\caption{The effects of anisotropic noise (generated by
$L = \kappa \sigma_{x}$ or $L = \kappa \sigma_{z}$)
on the gate Eq.(\ref{eq3}) are shown in a) for
$\omega_{0} - \omega = 100$. For adiabaticity $T$ takes 
(dimensionless) time $\pi$
and $C$ takes $2 \pi$. The $\Pi$'s are square pulses taking time $\pi /100$.
The ratio of these slow and fast operations is comparable to the 
frequency ratio in
\cite{jones1}. The loss in fidelity $(1 - f)$ is shown against the noise 
parameter
$\Gamma = \kappa^{2}$: $\gamma_{B} = \pi/8$, $z$-noise (i) and $x$-noise
(iv); $\gamma_{B} = \pi$, $z$-noise (ii) and $x$-noise (iii).
In b) the effects of isotropic noise (generated by three independent
$L_{i} = \kappa \sigma_{i}$ for $i=x,y,z$) on the gate Eq.(\ref{eq3}) for
$\gamma_{B} = \pi$. $(1 - f)$ (i)
and entropy $S = - Tr(\rho \log_{2} \rho)$ (ii)
are shown against $\Gamma = \kappa^{2}$.} \label{fig1}
\end{figure}
The case of isotropic noise is illustrated in Fig. 1(b). The small-$\Gamma$
rate of fidelity loss is twice that of the worst behaviour in Fig. 1(a) (which
follows from a simple analytic estimate) and indeed the whole fidelity loss
fits well with the analytic form
$1 - f = \frac{1}{2}(1 - \exp(-4 \Gamma \tau))$ where $\tau$ is the gate
duration. Provided that the gate is adiabatic, the effects of isotropic noise
are set by the gate length and level of decoherence, independent of the gate
details. Also shown in Fig. 1(b) is the final system entropy as a function of
$\Gamma$. For small $\Gamma$ the entropy (loss of information)
increases significantly faster than the loss in fidelity.

Our second example is the more important case of a conditional two-qubit
geometric gate \cite{ekert1}, where entanglement is generated, or not,
depending upon the level of decoherence. This requires two spin qubits
with bare transition frequencies of $\omega_{a}$ (target qubit) and
$\omega_{b}$ (control qubit) and $\omega_{a} \gg \omega_{b}$. An interaction
Hamiltonian of $H_{int} = \frac{J}{4} \sigma_{az} \sigma_{bz}$ generates the
conditional phase. This form of interaction is that appropriate for NMR and
certain condensed matter qubits. We have left $J$ fixed for the simulations
presented here (as is appropriate for NMR systems), but in principle this
coupling may be tunable for some condensed matter scenarios. A conditional
phase gate (introducing $\Delta \gamma = \gamma_{B}(\uparrow_{b}) -
\gamma_{B}(\downarrow_{b})$, for the two states of the control) is realised
by the unitary sequence $U_{\Delta \gamma} = \Pi^{b} U_{\gamma_{B}}^{a}
\Pi^{b} U_{\gamma_{B}}^{a}$. Here superscripts refer to the qubit operated
upon and $U_{\gamma_{B}}$ is the same as $U_{4 \gamma}$  but with the final
$\Pi$ removed. For $|\Psi\rangle_{ab} = |\psi\rangle_{a} |\psi\rangle_{b}$ this
generates
\begin{eqnarray} U_{\Delta \gamma} |\Psi\rangle_{ab} = \frac{1}{2}
\left(\exp(-2 i \Delta \gamma) |0\rangle_{a} |0\rangle_{b} + \exp(2 i \Delta
\gamma) |0\rangle_{a} |1\rangle_{b} \right. \nonumber \\ \left. + \exp(2 i
\Delta \gamma) |1\rangle_{a} |0\rangle_{b} + \exp(-2 i \Delta \gamma)
|1\rangle_{a} |1\rangle_{b}\right) \;. \label{eq4}
\end{eqnarray}
For a phase of $\Delta \gamma = \frac{\pi}{8}$ this state has a concurrence
\cite{conref} of unity and so is a maximally entangled two-qubit state.

We have chosen the coupling and frequency parameters so that the zero
decoherence state at the end of the simulated conditional phase gate is
maximally entangled with fidelity of 0.999946, and then investigated this gate
under various forms of decoherence. Examples of the results are shown in
Fig. 2.
\begin{figure}[!h]
\center{\epsfig{figure=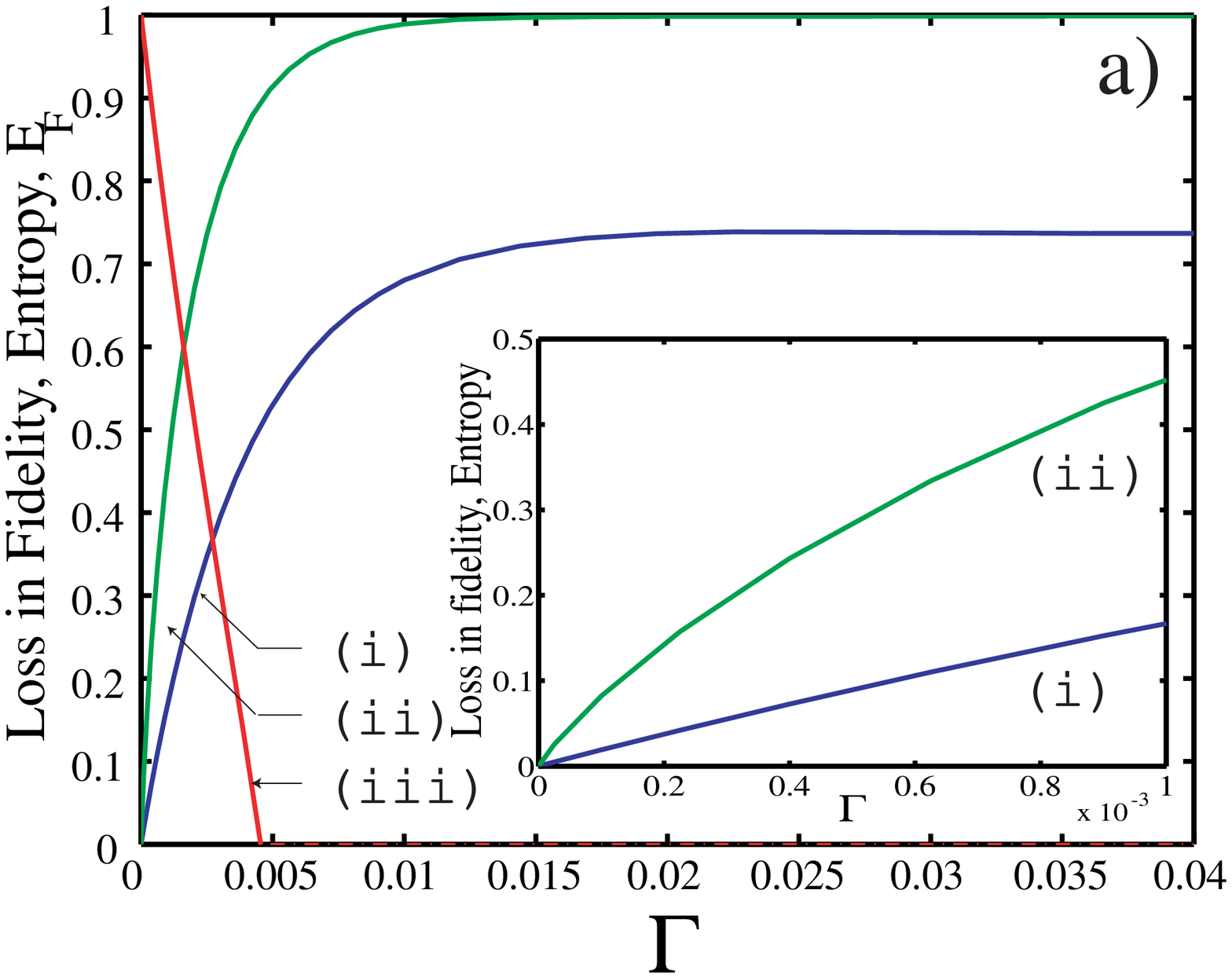,width=40mm}$\;\;\;$
\epsfig{figure=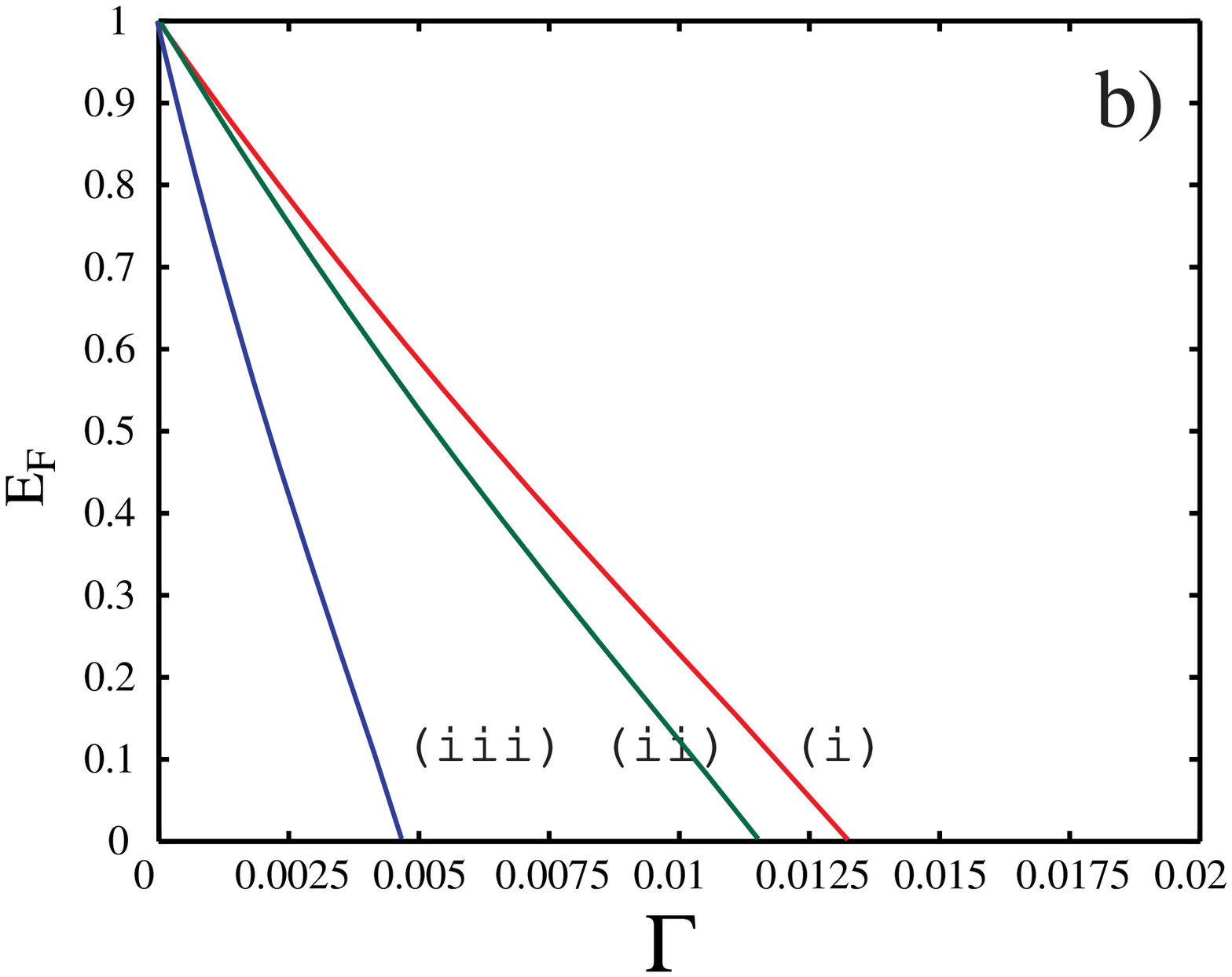,width=40mm}}
\caption{The effects of equal isotropic noise ($L_{i} = \kappa \sigma_{i}$
for $i=x,y,z$) applied to both qubits for the gate of Eq.(\ref{eq4})
are displayed in a) with $\Delta \gamma = \frac{\pi}{8}$.
Shown are plots of the loss in fidelity (i),
the entropy $S = - Tr(\rho \log_{4} \rho)$ (ii) and the entanglement of
formation (EOF) (iii) as a function of $\Gamma = \kappa^{2}$. 
The gate timings are
as in Fig. 1 and additional parameters used are $\omega_{a}-\omega = 100$,
$\omega_{b} = 1$, $\omega_{1} = 87.9238$ and $J = 37.5$. The entropy is base 4
to ensure a maximum value of 1. Also shown in a) is an
enlargement of the region near $\Gamma=0$. In b) we also show the decrease in
EOF as a function of $\Gamma$ for $s_x$ (i), $s_z$
(ii) and isotropic (iii) noise on each qubit.} \label{fig2}
\end{figure}
Clearly the rate
of fidelity loss and the sympathetic increase in entropy are correspondingly
greater for this system as noise is acting independently on the two qubits. For
this system $\rho$ was reconstructed by tomography \cite{tomogref} from the
sixteen expectation values $\langle \sigma_{ai} \sigma_{bj} \rangle$ for
$i,j=0,x,y,z$ ($\sigma_{0} \equiv Identity$), akin to what is needed in any
two-qubit experiment for a full reconstruction of $\rho$. From this it is
possible to compute the entropy and some measure of entanglement. For
illustration we use the entanglement of formation (EOF) \cite{eofref} as this
gives an upper bound on the level of decoherence 
for which some entanglement can be said to exist 
($\Gamma_{thres} = 0.00445$ for isotropic noise in our example).

The maximally entangling parameters used for Fig. 2 generate relatively large 
tipping angles, so there is
only a slight sensitivity to the direction of anisotropic noise applied to
both qubits, as shown in Fig. 2(b). Furthermore, detailed studies not 
illustrated here show that there is only a minor 
difference between the separate
effects of noise on the control and the target, so from an experimental
perspective on such a gate, there is nothing to be gained by singling out
either of these for decoherence reduction measures (e.g. error correction). 
Both these points hold right down into the very small decoherence regime,
where any practical system would have to operate.

In our simulations so far the gate times have been set to ensure
adiabaticity. Such gates are relatively slow 
and therefore
exposed to the ravages of decoherence for longer. Conventional dynamic gates 
can run much more quickly and so for comparison we investigate a dynamic
gate based on the same interaction as the entangling adiabatic gate. 
The gate is realised by the
unitary evolution $U(T) = \exp \left[ i\frac{J T}{4} \sigma_{az}
\sigma_{bz}\right]$ where we again choose $J=37.5$ and 
$T$ is now the interaction
time. In the absence of decoherence this produces the maximally entangled
state $|\Psi\rangle_{ab} = 
\frac{1}{2} e^{-i {\pi\over 4}} \left(  |0\rangle_{a}
|0\rangle_{b} + i|0\rangle_{a} |1\rangle_{b} + i |1\rangle_{a} |0\rangle_{b} +
|1\rangle_{a} |1\rangle_{b}\right)$ with a fidelity of 1 if the
gate acts for a total time $T_{tot}=2 \pi /75$.
For the adiabatic geometric phase gate $T_{tot}=12.0004 \pi$,
so the dynamic gate is
approximately 450 times faster and its speed is limited directly by the
strength of $J$. The results for this dynamic gate are illustrated in 
Fig. 3(a).
\begin{figure}[!h]
\center{\epsfig{figure=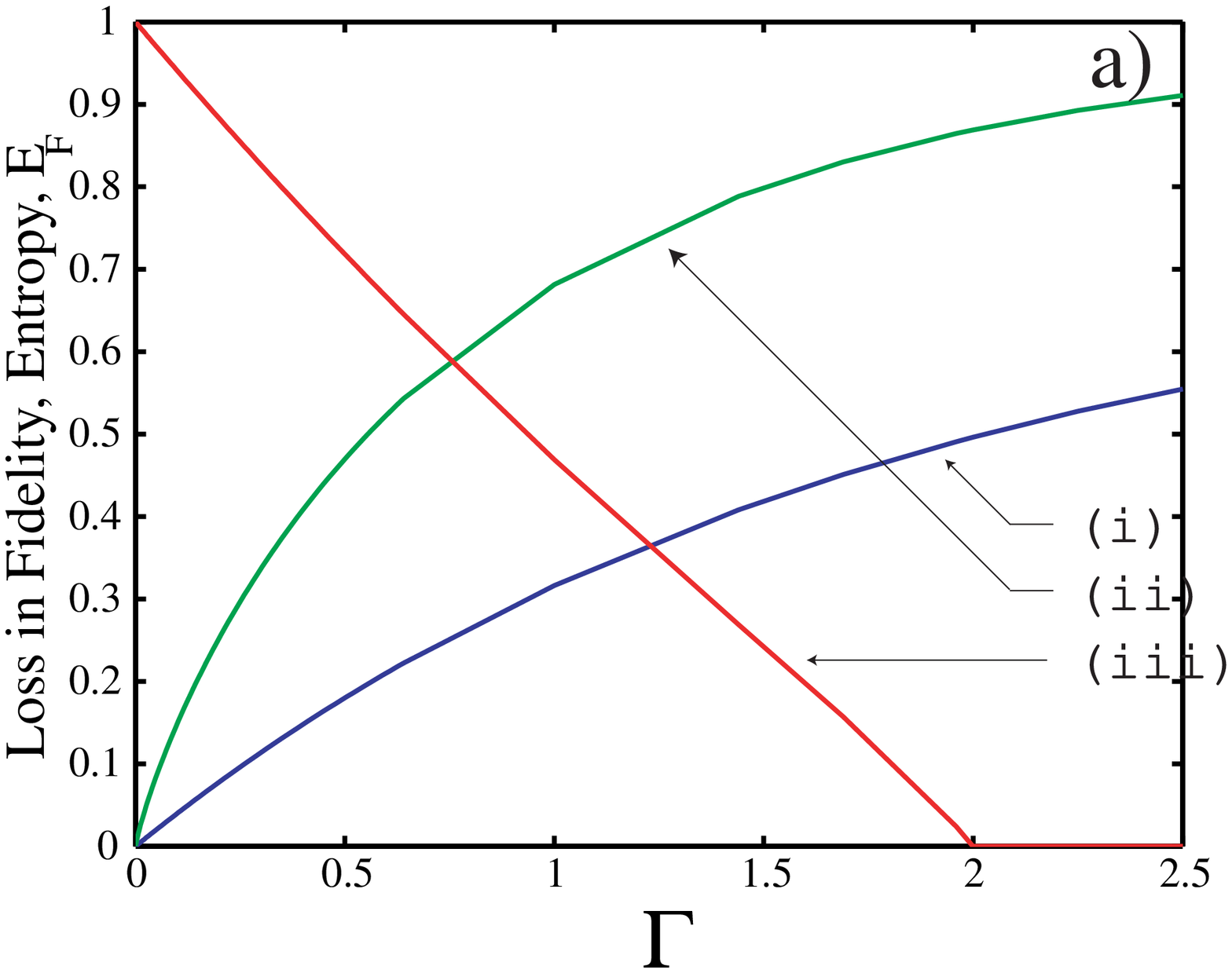,width=40mm}$\;\;\;$
\epsfig{figure=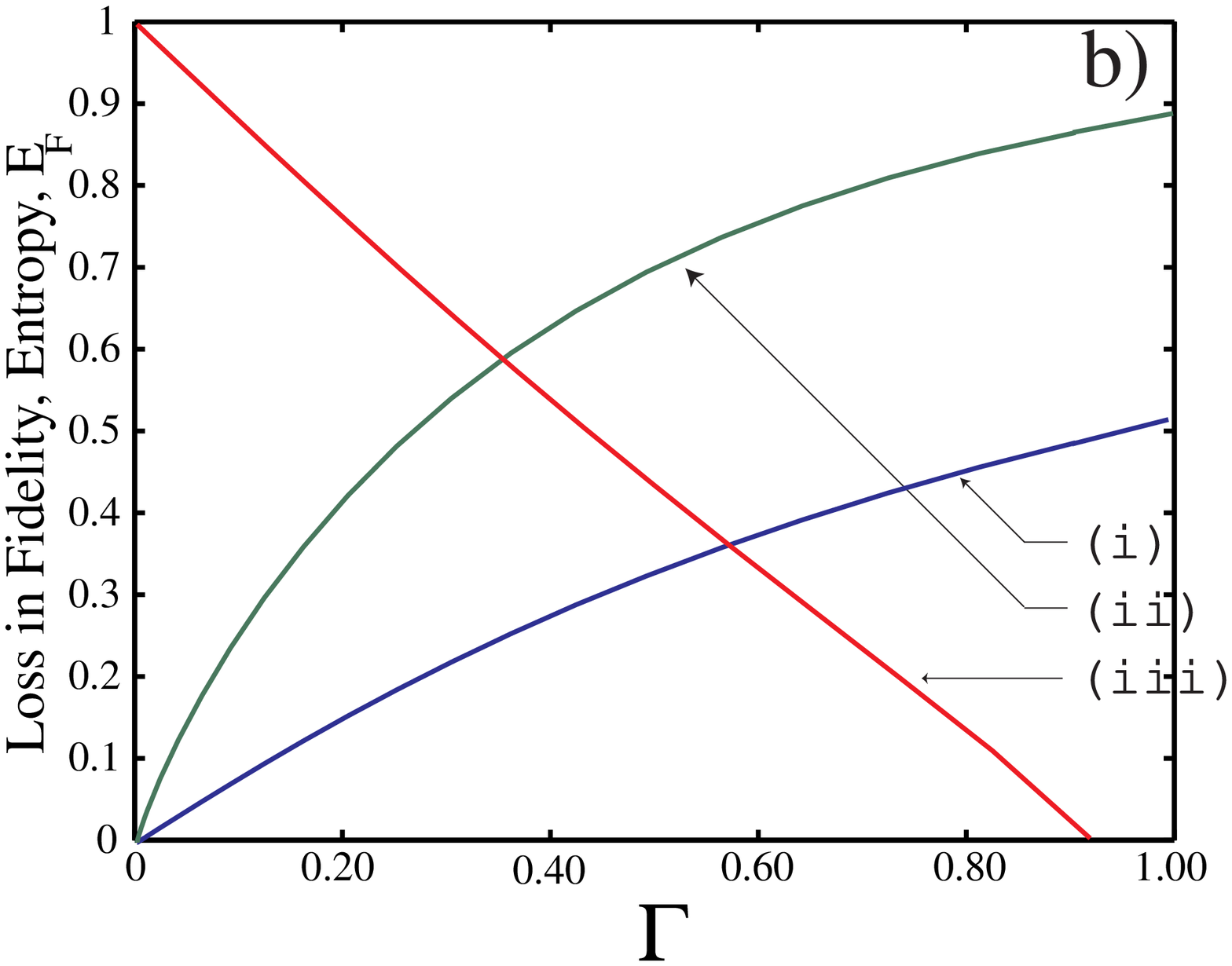,width=40mm}} \caption{The 
effects of equal isotropic
noise  applied to both qubits for the
dynamic gate  are displayed in a) and to the nonadiabatic geometric gate in b).
Shown in the  plots are the loss in fidelity (i), the entropy $S = - Tr(\rho
\log_{4} \rho)$ (ii) and the entanglement of formation (iii) as a function of
$\Gamma$. For the dynamic gate the following parameters were chosen:
$J=37.5$ and
$T=\pi /J$. For the nonadiabatic fast geometric gate $J=37.5$, $\delta
\omega=18.75$, $\omega_B=0.01$.}
\label{fig3}
\end{figure}
As isotropic noise acts on the dynamic gate, the entanglement of
formation falls to zero at $\Gamma_{thres} \sim 2$ (compared with
$\Gamma_{thres} = 0.00445$ for the adiabatic gate). In fact we find that
$\Gamma_{thres}\; T_{tot}\sim 4 \pi /75$ for both the adiabatic geometric gate
and the dynamic gate. Therefore, as the adiabatic gate operates for a 
significantly
longer time, it is much more severely affected by decoherence. 
This has serious 
implications
for the physical realisation of such a geometric gate. 
Recently, however, it has
been proposed to use the non-adiabatic, or Aharonov-Anandan, phase to speed up
geometric phase gates \cite{xbin1}. Here the achievable reduction in 
decoherence
is not entirely clear as this technique also introduces new potential
decoherence sources \cite{blais1}. We implement the unitary sequence
$R_x^a(\pi/2)R_z^a(-\pi/2)\tilde U(\pi/4)R_y^a(-\pi/2)$ where
$\tilde U(\pi/4)=R_x^a(3\pi/4)U(\pi/J) R_x^a(-\pi/2)U(\pi/J) R_x^a(-\pi/4)$
to produce the maximally entangled state $|\Psi\rangle_{ab}$.
Here $R_x^a (\theta)$ indicates a rotation of $\theta$ about the $x$ axis.
The results of isotropic noise acting on both qubits are displayed in 
Fig. 3(b). As
this fast geometric gate has two periods of free evolution it acts 
for twice as long as the dynamic gate.
Hence, it is subject to the effects of decoherence for longer and 
entanglement is again lost at a faster rate (slightly greater than twice the 
rate, $\Gamma_{thres} \sim 0.945$). Overall, our results suggest that
geometric gates probably only offer a real advantage if they can be
implemented faster than the equivalent dynamical gates. 
At present, such proposed gates do not beat the dynamic gate time 
($\pi/J$ for our example), and it is not obvious that this is possible given
both approaches realise entanglement through interacting qubit evolution. 
However, whether non-adiabatic geometric gates can be 
implemented more quickly is
an open and critically important question currently under investigation.


A number of comments can be made in conclusion.\\
(1) Single qubit geometric phase gates can show some sensitivity to
the direction of anisotropic noise, but this is path-dependent. In our
example, if the path is chosen to generate a large ($\sim \pi$) relative
phase between the qubit amplitudes, there is very little sensitivity.\\
(2) From the general perspective of our simulation approach, single qubit
gate behaviour can be used with experimental results to calibrate the level 
of decoherence present in a system\footnote{All the quantities used in
our simulations are dimensionless---an experimentally measured quantity such as
a transition frequency sets the scale for everything in any actual qubit
realisation.}. Although the anisotropic noise-sensitive
gates may be of limited use for actual quantum information processing (due to
the small phase difference generated), they may be applied, for
example coupled with the ability to reorientate the external 
static magnetic field ($z$-axis), in mapping out the {\em forms} of decoherence
acting on a qubit, in addition to calibrating them. This could be extremely
useful for new experimental systems where the dominant environmental coupling
is unclear in advance.\\
(3) The single qubit decoherence calibrations can be used to predict the
expected level of entanglement in two-qubit gates (as illustrated in Fig. 2)
prior to experiment.\\
(4) The adiabatic gates discussed here are
slow (relative to the timescale for dynamic gates) and so
exposed to the ravages of decoherence for longer. To overcome this it is
necessary to perform geometric gates faster than the equivalent dynamic
gate. Whether they
can be made faster than the dynamic gate is a unanswered question left for
future investigation. The practical use of such geometric gates depends upon 
the resulting answer.


We thank Jonathan Jones, Andrew Briggs and R\"{u}diger Schack for helpful
conversations and Richard Cardwell for technical assistance. Work supported
in part by European Commission grant IST-1999-29110 MAGQIP.

\end{document}